\documentclass[12pt]{article}

\usepackage{amsmath}
\usepackage{amsfonts}

\newtheorem{theorem}{Theorem}
\newtheorem{lemma}[theorem]{Lemma}

\newtheorem{definition}[theorem]{Definition}

\begin{document}

\title{Multidimensional basis of $p$-adic wavelets \\ and representation theory}

\author{S.Albeverio\footnote{Universit\"at Bonn,
Institut f\"ur Angewandte Mathematik, Bonn, Germany}, 
S.V.Kozyrev\footnote{Steklov Mathematical Institute, Moscow, Russia}}

\maketitle

\begin{abstract}
A multidimensional basis of $p$-adic wavelets is
constructed. The relation of the constructed basis to a system of
coherent states (i.e. orbit of action) for some $p$-adic group of linear
transformations is discussed. We show that the set of products of
the vectors from the constructed basis and $p$-roots of one is the orbit of the corresponding $p$-adic group of
linear transformations.
\end{abstract}


\bigskip


\section{Introduction}

In the present paper we construct a multidimensional $p$-adic wavelet
basis and describe the relation of this basis to representation
theory, namely to the theory of coherent states for groups of linear
transformations.

The first basis of $p$-adic wavelets (in one dimension) was constructed in
\cite{wavelets}. An analogous basis (with generalizations to some
abelian groups) was built in \cite{Benedetto}. In \cite{ShelkSkop},
\cite{KhrShelkSkop} some other examples of $p$-adic wavelet bases
were proposed.

An example of multidimensional $p$-adic wavelet basis for $p=2$ was constructed, with the help of the $p$-adic multiresolution construction, in \cite{ShelkSkopina}. This example coincides (for $p=2$) with the considered in the present paper. We show that this multiresolution wavelet basis can be described by the simple formula (\ref{basis_intro}) below.

In \cite{frames} it was shown that the orbit of the action of the $p$-adic
affine group (i.e. the system of coherent states for this group) on
a generic function from the space $D_0(\mathbb{Q}_p)$ (of mean zero
locally constant compactly supported complex valued functions of
$p$-adic argument) gives a tight uniform frame of wavelets. An
explicit parametrization for this frame was obtained and the bound
was computed.

In the simplest case (when we consider the orbit for the wavelet
$\psi(x)=\chi(p^{-1}x)\Omega(|x|_p)$, see \cite{cont_wavelets}) the
frame of wavelets discussed above contains the products of wavelets
from the $p$-adic wavelet basis and $p$-roots of one.

In the present paper we construct a basis of
multidimensional $p$-adic wavelets and investigate the relation of
this basis with the structure of orbit of $p$-adic group of
transformations. The constructed basis is the direct generalization
of the one dimensional $p$-adic wavelet basis of \cite{wavelets}. In particular
wavelets with the support in the unit $d$-dimensional ball are
introduced by the formula
$$
\psi_J(x)=\chi(p^{-1}Jx)\Omega(|x|_p),
$$
$x\in \mathbb{Q}_p^d$, $|\cdot|_p$ is the $d$-dimensional $p$-adic
norm, $J$ is the set of representatives from the maximal subballs of a
$p$-adic $d$-dimensional sphere, $Jx$ is the scalar product in
$\mathbb{Q}_p^d$.

We define the multidimensional basis of $p$-adic wavelets by the formula
\begin{equation}\label{basis_intro}
\psi_{\gamma n J}(x)=p^{-{d\gamma\over
2}}\psi_{J}(p^{\gamma}x-n),\qquad x\in \mathbb{Q}_p^d,\quad
\gamma\in\mathbb{Z},\quad n\in \mathbb{Q}_p^d/\mathbb{Z}_p^d.
\end{equation}
This formula is a direct generalization of the construction of the $p$-adic wavelet basis in one dimension \cite{wavelets}.

We show that the frame of wavelets obtained by multiplication of $\psi_{\gamma n J}$ by $p$-roots of one is the orbit of the
group of transformations generated by translations, uniform
dilations of all coordinates and by linear transformations which
conserve the $d$-dimensional norm.

This result is the example of the {\it orbital approach } to $p$-adic wavelets, proposed in \cite{frames}: wavelet frames should be considered as orbits of the corresponding groups of transformations and $p$-adic wavelet analysis in general is a part of the representation theory for some $p$-adic groups of transformations.

The exposition of the present paper is as follows.

In Section 2 we recall the earlier constructions of the
basis of $p$-adic wavelets and of the relation between the frames of
$p$-adic wavelets and the theory of coherent states for the affine
group.

In Section 3 we introduce the basis of multidimensional $p$-adic
wavelets.

In Section 4 we compare the constructed basis with the multiresolution approach to wavelets.

In Section 5 we show that the introduced multidimensional $p$-adic
wavelet basis can be considered (up to multiplication by $p$-roots of one) as the orbit of some group of linear transformations and translations.

\section{One--dimensional $p$-adic wavelets}

In the present Section we recall the construction of the basis of
$p$-adic wavelets and the relation of the frames of $p$-adic wavelets  with the orbits of the
affine group.

Consider a set of wavelets related to the unit ball in the field $\mathbb{Q}_p$ of $p$-adic numbers in the form of
the following complex valued functions of a $p$-adic argument
\begin{equation}\label{psi}
\psi_k(x)=\chi(p^{-1}kx)\Omega(|x|_p),\qquad x,k\in \mathbb{Q}_p,
\end{equation}
where $|k|_p=1$.

Here $\Omega$ is the characteristic function of the interval $[0,1]$ (i.e. $\Omega(|x|_p)$ is the characteristic function of the unit ball with the center in zero in $\mathbb{Q}_p$), and $\chi$ is the complex valued character of the $p$-adic argument:
$$
\chi(x)=e^{2\pi i\{x\}},
$$
where $\{x\}$ is the fractional part of $x$:
$$
\{x\}=\sum_{j=\gamma}^{-1}x_jp^{j},\quad x_j=0,\dots,p-1
$$
for the expansion of the $p$-adic number $x$ over degrees of $p$:
$$
x=\sum_{j=\gamma}^{\infty}x_jp^{j},\quad x_j=0,\dots,p-1.
$$

We have exactly $p-1$ different functions of the form (\ref{psi}) (considered
as functions of $x$) due to the fact that $\psi_k(x)$ is locally constant as a
function of $k$.

Namely taking the representatives $j=1,\dots,p-1$ in the maximal
subballs of the sphere $|k|_p=1$ we get the set of functions
$$
\psi_{j}(x)=\chi(p^{-1}jx)\Omega(|x|_p),\qquad j=1,\dots,p-1.
$$

All the above functions are dilations of the $p$-adic wavelet
$\psi(x)=\chi(p^{-1}x)\Omega(|x|_p)$:
$$
\psi_{j}(x)=\psi(jx).
$$
Therefore the orbit of the group of dilations from the unit sphere
acting on the wavelet $\psi$ is exactly the set of wavelets
$\psi_j$.

Then we construct \cite{wavelets} the basis $\{\psi_{\gamma n j}\}$ of wavelets by
application to the set of functions $\{\psi_{j}\}$ of dilations to
degrees of $p$ and translations by the representatives of the
equivalence classes from the factor group
$\mathbb{Q}_p/\mathbb{Z}_p$:
\begin{equation}\label{thebasis}
\psi_{\gamma n j}(x)=p^{-{\gamma\over
2}}\psi_j(p^{\gamma}x-n),\qquad x\in \mathbb{Q}_p,\quad
\gamma\in\mathbb{Z},\quad n\in \mathbb{Q}_p/\mathbb{Z}_p,
\end{equation}
\begin{equation}\label{QpZp}
n=\sum_{i=\beta}^{-1}n_{i}p^{i},\quad n_i=0,\dots,p-1.
\end{equation}

The affine group acts in $L^2(\mathbb{Q}_p)$ by translations and
dilations
$$
G(a,b)f(x)=|a|_p^{-{1\over 2}}f\left({x-b\over a}\right),\qquad
a,b\in\mathbb{Q}_p,\quad a\ne 0.
$$

The orbit of the action of the affine group on the wavelet $\psi(x)=\chi(p^{-1}x)\Omega(|x|_p)$ (i.e.
the system of coherent states for this group) will be a frame of
wavelets which contains all the products of wavelets from the above basis
$\{\psi_{\gamma n j}\}$ of $p$-adic wavelets and the $p$ different $p$-roots of one, i.e. $e^{2\pi i p^{-1}l}$, $l=0,1,\dots,p-1$.

In the general case, it was found that \cite{frames} for a generic function $f$ (see \cite{frames} for the definitions) belonging to the space
$D_0(\mathbb{Q}_p)$ of complex valued mean zero locally constant
compactly supported functions of a $p$-adic argument we have the
following theorem: the orbit of this function with respect to action of the affine group is a uniform and tight frame in $L^2(\mathbb{Q}_p)$, and the bound of the frame can be computed explicitly.

Moreover, this orbit possesses the explicit parametrization. The orbit is parameterized by the three indices: the index $\gamma$ describes the dilations by the degrees of $p$, the index $n$ describes the translations by elements of $\mathbb{Q}_p/\mathbb{Z}_p$, and the index $J$ takes the finite number of values. Therefore the parameters (translations $n$ and dilations $p^{\gamma}$) which are postulated
for the constructions of bases and frames of
wavelets arise as parameters on
the orbit of a generic function from $D_0(\mathbb{Q}_p)$. Therefore
in the $p$-adic case the multiresolution wavelet analysis is a particular
case of the structure of the orbit of the function $f\in
D_0(\mathbb{Q}_p)$ with respect to the action of the affine group.

In general, the theory of $p$-adic wavelets can be considered as a part of the representation theory
for $p$-adic groups of transformations.
One can consider orbits of action for the different groups of
$p$-adic linear transformations (i.e. the systems of coherent states
for the different groups, see \cite{Perelomov}) in the spaces of functions of a
multidimensional $p$-adic argument. In the present paper we
investigate two examples of multidimensional $p$-adic wavelet
frames related to systems of coherent states.

\section{Multidimensional $p$-adic wavelet basis}

In the present Section we introduce our multidimensional basis of
$p$-adic wavelets. Let us consider (analogously to the one
dimensional case) the set of functions
\begin{equation}\label{basis_0}
\psi_k(x)=\chi(p^{-1}kx)\Omega(|x|_p),\qquad x,k\in \mathbb{Q}^d_p,
\end{equation}
where $|k|_p=1$ and
$$
kx=\sum_{l=1}^d k_l x_l.
$$

Let us remind that the $d$-dimensional $p$-adic norm is introduced
as
$$
|x|_p=\max_{l=1,\dots,d} |x_l|_p.
$$
Thus the $d$-dimensional $p$-adic ball is a direct product of $d$
one dimensional balls, i.e. it coincides with the $d$-dimensional
cube. The $d$-dimensional sphere is the $d$-dimensional ball without the
subball with the diameter which is $p$ times less than the diameter
of the initial ball.

There are $p^d-1$ different functions $\psi_k(x)$ (as functions of
$x$). We choose the following representatives $J$ for the
$d$-dimensional $k$, $|k|_p=1$, which enumerate these functions:
\begin{equation}\label{basis_01}
k=J=\left(j_1,\dots,j_d\right),\qquad j_l=0,\dots,p-1,
\end{equation}
where at least one of $j_l$ is not equal to zero. This set of
$J$ is the set of representatives from the maximal ($d$-dimensional)
subballs of the $p$-adic $d$-dimensional sphere.

We build the basis $\{\psi_{\gamma n J}\}$ of $d$-dimensional
wavelets by application to the set of functions $\{\psi_{J}\}$ of
dilations by the degrees of $p$ and translations by the
representatives of the equivalence classes of the factor group
$\mathbb{Q}_p^d/\mathbb{Z}_p^d$:
\begin{equation}\label{basis_1}
\psi_{\gamma n J}(x)=p^{-{d\gamma\over
2}}\psi_{J}(p^{\gamma}x-n),\qquad x\in \mathbb{Q}_p^d,\quad
\gamma\in\mathbb{Z},\quad n\in \mathbb{Q}_p^d/\mathbb{Z}_p^d,
\end{equation}
\begin{equation}\label{basis_2}
n=\left(n^{(1)},\dots,n^{(d)}\right),\qquad
n^{(l)}=\sum_{i=\beta_l}^{-1}n^{(l)}_{i}p^{i},\quad
n_i^{(l)}=0,\dots,p-1,\quad \beta_l\in\mathbb{Z}_{-}.
\end{equation}
Here $\mathbb{Z}_{-}$ is the set of negative integers.

\begin{theorem}\label{d-basis}\quad{\sl
The set of functions $\{\psi_{\gamma n J}\}$ defined by
(\ref{basis_1}), (\ref{basis_2}) is an orthonormal basis in
$L^2(\mathbb{Q}_p^d)$. }
\end{theorem}

\noindent{\it Proof}\qquad The proof is analogous to the proof of the corresponding theorem in the one
dimensional case \cite{wavelets}. It is easy to see that the wavelet
is a mean zero function:
\begin{equation}
\label{intzero} \int_{\mathbb{Q}_p^d}\psi_{\gamma nJ}(x)\,d\mu(x)=0.
\end{equation}

The product of wavelets $\psi_{\gamma nJ}\psi_{\gamma' n'J'}$, where
$\gamma<\gamma'$, is the wavelet $\psi_{\gamma nJ}$ multiplied by a
number, since $\psi_{\gamma' n'J'}$ is a constant on the support of
$\psi_{\gamma nJ}$. This and (\ref{intzero}) imply that
$$
\langle \psi_{\gamma nJ},\psi_{\gamma' n'J'}\rangle=0,\qquad
\gamma\ne \gamma'.
$$

Let us multiply the characteristic functions of the supports of
wavelets $\psi_{\gamma nJ}$, $\psi_{\gamma' n'J'}$. For
$\gamma\le\gamma'$ the following product is equal to the
characteristic function or zero:
$$
\Omega\bigl(|p^{\gamma}x-n|_p\bigr)
\Omega\bigl(|p^{\gamma'}x-n'|_p\bigr)=
\Omega\bigl(|p^{\gamma}x-n|_p\bigr)
\Omega\bigl(|p^{\gamma'-\gamma}n-n'|_p\bigr).
$$
For $\gamma=\gamma'$ and $n$, $n'\in \mathbb{Q}^d_p/\mathbb{Z}^d_p$,
we get at the RHS of the above equation
$$
\Omega(|n-n'|_p)=\delta_{nn'}.
$$
This implies that the scalar product of wavelets $\psi_{\gamma nJ}$,
$\psi_{\gamma n'J'}$ can be non zero only for $n=n'$.

Computing
\begin{align*}
\langle \psi_{\gamma nJ},\psi_{\gamma' n'J'}\rangle
&=\delta_{\gamma\gamma'}\delta_{nn'} \int_{\mathbb{Q}_p^d}
p^{-\gamma} \chi(p^{\gamma-1}(J'-J)(x-p^{-\gamma}n)) \times
\\
&\quad\times \Omega(|p^{\gamma}x-n|_p)\,d\mu(x) =
\delta_{\gamma\gamma'}\delta_{nn'}\delta_{JJ'},
\end{align*}
we prove the orthonormality of the vectors $\psi_{\gamma nj}$.

To prove that the set of vectors $\{\psi_{\gamma nj}\}$ is an
orthonormal basis in $L^2(\mathbb{Q}^d_p)$ (i.e. to prove the
completeness of this set) we use the Parseval identity.

Since the set of characteristic functions of balls is complete in
~$L^2(\mathbb{Q}^d_p)$, and the action of the group of translations $\mathbb{Q}^d_p/\mathbb{Z}^d_p$ and dilations by $p^{\gamma}$, $\gamma\in\mathbb{Z}$ is transitive on the set of balls in $\mathbb{Q}^d_p$ (see lemma \ref{ball_orbit} below), it is sufficient to prove the Parseval
identity for the characteristic function $\Omega(|x|_p)$. We get
\begin{equation}
\label{completeness0}
\begin{gathered}
\langle\Omega(|x|_p),\psi_{\gamma nJ}\rangle= p^{-{d\gamma\over
2}}\theta(\gamma)\delta_{n0},
\\
\theta(\gamma)=0,\quad \gamma\le 0,\qquad \theta(\gamma)=1,\quad
\gamma\ge 1.
\end{gathered}
\end{equation}
This implies the Parseval identity for $\Omega(|x|_p)$:
$$
\sum_{\gamma nJ}|\langle\Omega(|x|_p),\psi_{\gamma nJ}\rangle|^2
=\sum_{\gamma=1}^{\infty}(p^d-1)p^{-d\gamma}=1
=|\langle\Omega(|x|_p),\Omega(|x|_p)\rangle|^2.
$$

This finishes the proof of the theorem. $\square$

\section{Comparison with the multiresolution construction}

There exists the standard way of constructing of multidimensional wavelet bases from one dimensional bases using the multiresolution construction. In the $p$-adic case (for $p=2$) the corresponding example of a multidimensional wavelet basis was considered in \cite{ShelkSkopina}.

The following definition of the $p$-adic multiresolution analysis \cite{cont_wavelets}, \cite{ShelkSkopina}, is the direct analogue of the real multiresolution construction, cf. \cite{Meyer}, \cite{Daubechies}.

\begin{definition}\label{MRA}\qquad{\sl
A system of closed subspaces $V_{\gamma}\subset L^2(\mathbb{Q}_p)$, $\gamma\in\mathbb{Z}$, is called a
multiresolution analysis (MRA) in $L^2(\mathbb{Q}_p)$ if the following properties are satisfied:

(i) $V_{\gamma}\subset V_{\gamma+1}$ for all $\gamma\in\mathbb{Z}$;

(ii) $\bigcup_{\gamma\in\mathbb{Z}}V_{\gamma}$ is dense in $L^2(\mathbb{Q}_p)$;

(iii) $\bigcap_{\gamma\in\mathbb{Z}}V_{\gamma}=\{0\}$;

(iv) $f(x)\in V_{\gamma} \Longleftrightarrow f(p^{-1}x)\in V_{\gamma+1}$ for all $\gamma\in\mathbb{Z}$;

(v) there exists a function $\phi\in V_0$ such that the system $\{\phi(x-n)\}$, $n\in\mathbb{Q}_p/\mathbb{Z}_p$, is
an orthonormal basis in $V_0$.
}
\end{definition}

Here we as usual take $n\in\mathbb{Q}_p/\mathbb{Z}_p$ in the form of fractions (\ref{QpZp}). The function $\phi$ above is called the scaling (refinable) function. The system $\{\phi(p^{\gamma}x-n)\}$, $n\in\mathbb{Q}_p/\mathbb{Z}_p$, will be an orthonormal basis in $V_{-\gamma}$.

Then we define the wavelet spaces $W_{\gamma}$ as the orthogonal complements to $V_{\gamma}$ in $V_{\gamma+1}$:
$$
V_{\gamma+1}=V_{\gamma}\oplus W_{\gamma}.
$$
We have
$$
L^2(\mathbb{Q}_p)=\oplus_{\gamma\in\mathbb{Z}} W_{\gamma},
$$
(here $\oplus$ is the completion of the direct sum) and for $\gamma\in\mathbb{Z}$
$$
f(x)\in W_{\gamma} \Longleftrightarrow f(p^{-1}x)\in W_{\gamma+1}.
$$

Then, taking a final set of functions (wavelets) $\psi_j\in W_0$ such that $\{\psi_j(x-n)\}$, $n\in\mathbb{Q}_p/\mathbb{Z}_p$, is
an orthonormal basis in $W_0$, we construct the basis of multiresolution wavelets in $L^2(\mathbb{Q}_p)$ of the form  $\{p^{-{\gamma\over 2}}\psi_j(p^{\gamma}x-n)\}$, $n\in\mathbb{Q}_p/\mathbb{Z}_p$, $\gamma\in\mathbb{Z}$.

\bigskip

\noindent{\bf Example 1}\qquad
The $p$-adic wavelet basis (\ref{thebasis}) is obtained in the MRA approach by taking
$$
\phi(x)=\Omega(|x|_p),\qquad \psi_j(x)=\chi(p^{-1}jx)\Omega(|x|_p),\quad j=1,\dots,p-1.
$$

\bigskip

The multidimensional multiresolution analysis in $L^2(\mathbb{Q}^d_p)$ is introduced by taking the system of tensor product of one dimensional subspaces $V_{\gamma}^{(l)}$:
$$
V_{\gamma}=\otimes_{l=1}^d V_{\gamma}^{(l)},\quad \gamma\in\mathbb{Z}.
$$
The scaling function is the tensor product of the one dimensional scaling functions
$$
\Phi=\otimes_{l=1}^d \phi^{(l)}.
$$
The translations  $\{\Phi(p^{\gamma}x-n)\}$, $n\in\mathbb{Q}^d_p/\mathbb{Z}^d_p$, will be an orthonormal basis in $V_{-\gamma}$.

Therefore the system of spaces $V_{\gamma}$ and the scaling function $\Phi$ satisfy the multidimensional generalization of definition \ref{MRA}.

Then we define the wavelet spaces $W_{\gamma}$ as orthogonal complements to $V_{\gamma}$ in $V_{\gamma+1}$:
$$
V_{\gamma+1}=V_{\gamma}\oplus W_{\gamma},
$$
$$
L^2(\mathbb{Q}^d_p)=\oplus_{\gamma\in\mathbb{Z}} W_{\gamma}.
$$

The multidimensional wavelet spaces have the form of the following direct sums of tensor products of one dimensional wavelet spaces
$$
W_{\gamma}=\oplus_{\epsilon_1,\dots,\epsilon_d: \epsilon_k=0,1\backslash \{0,\dots,0\}}\otimes_{l=1}^d V_{\gamma}^{(\epsilon_l)},
$$
where $V_{\gamma}^{(0)}=V_{\gamma}$, $V_{\gamma}^{(1)}=W_{\gamma}$, and we take the direct summation of all tensor products
of $W_{\gamma}$ and $V_{\gamma}$ where not all subspaces are taken to be $V_{\gamma}$ (i.e. not all indices $\epsilon$ are equal to zero).

The multidimensional multiresolution wavelets $\Psi_J$, $J=(j_1,\dots,j_d)$ are defined by the prescription
\begin{equation}\label{multi_prod}
\Psi_J=\otimes_{l=1}^d \psi_{j_l},
\end{equation}
where $\psi_j$ for $j\ne 0$ are equal to one dimensional wavelets $\psi_j$ or equal to $\phi$ for $j=0$.
Here we exclude from the set $\{\psi_J\}$  the product $\otimes_{l=1}^d \phi$ which corresponds to $J=(0,\dots,0)$.

The basis of multiresolution wavelets in $L^2(\mathbb{Q}^d_p)$ is introduced as the set of vectors  $\{p^{-{d\gamma\over 2}}\Psi_J(p^{\gamma}x-n)\}$, $n\in\mathbb{Q}^d_p/\mathbb{Z}^d_p$, $\gamma\in\mathbb{Z}$ given by translations and dilations of the finite set of the wavelets $\psi_J$.

\bigskip

\noindent{\bf Example 2}\qquad
The $p$-adic multidimensional wavelet basis (\ref{basis_1}) is obtained from the above construction in the following way.
Since in the dimension one we have
$$
\phi(x)=\Omega(|x|_p),\qquad \psi_j=\chi(p^{-1}jx)\Omega(|x|_p),\quad j=1,\dots,p-1,
$$
then, taking the tensor product (\ref{multi_prod}) of one dimensional wavelets
$$
\Psi_J=\otimes_{l=1}^d \psi_{j_l},\qquad J=\left(j_1,\dots,j_d\right),\qquad j_l=0,\dots,p-1,
$$
with $J\ne (0,\dots,0)$ we get
$$
\Psi_J(x)=\chi(p^{-1}Jx)\Omega(|x|_p),\qquad x\in\mathbb{Q}_p^d,\quad J=\left(j_1,\dots,j_d\right),\quad Jx=\sum_{l=1}^d J_lx_l,
$$
which coincides with the wavelets (\ref{basis_1}).
Therefore the simple prescription (\ref{basis_1}) for $\psi_J$ reproduces the more complicated multiresolution construction.

\section{Representation theory and frames of wavelets}

In the present Section we show the relation of the multidimensional wavelet basis (\ref{basis_1}), (\ref{basis_2})  of theorem \ref{d-basis} to the representation theory for some $p$-adic groups of transformations. For representation theory of $p$-adic groups see \cite{Gelfand}. 

The proof of the following lemma is straightforward.

\begin{lemma}\label{ball_orbit}\qquad
{\sl 1) The group $\mathbb{Q}_p^d/\mathbb{Z}_p^d$ of the following lines of fractions with addition modulo one
$$
n=\left(n^{(1)},\dots,n^{(d)}\right),\qquad
n^{(l)}=\sum_{i=\beta_l}^{-1}n^{(l)}_{i}p^{i},\quad
n_i^{(l)}=0,\dots,p-1,\quad \beta_l\in\mathbb{Z}_{-},
$$
acts transitively by translations on the set of all balls with the diameter one in $\mathbb{Q}_p^d$.

2) The group generated by the above translations from $\mathbb{Q}_p^d/\mathbb{Z}_p^d$ and dilations
$$
x\mapsto p^{\gamma}x, \quad\gamma\in\mathbb{Z}, \quad x\in\mathbb{Q}_p^d,
$$
acts transitively on the set of all balls in $\mathbb{Q}_p^d$.

3) A characteristic function of any ball in $\mathbb{Q}_p^d$ can be uniquely represented in the form
$$
\Omega(|p^{\gamma }x-n|),\qquad x\in\mathbb{Q}_p^d,\quad n\in\mathbb{Q}_p^d/\mathbb{Z}_p^d,\quad \gamma\in\mathbb{Z}.
$$

}
\end{lemma}

\begin{definition}\qquad
{\sl The group $O_d$ is the group of all linear transformations in
$\mathbb{Q}_p^d$ which conserve the $p$-adic norm:
$$
x\mapsto gx,\qquad (gx)_i=\sum_{j=1}^d g_{ij}x_j.
$$
Here $g\in O_d$, $x\in\mathbb{Q}_p^d$, $|gx|_p=|x|_p$.}
\end{definition}

This group can be considered as the $p$-adic analogue of the group of orthogonal transformations in $\mathbb{R}^d$.

Let us denote ${\cal V}$ the ball with the diameter one in $\mathbb{Q}_p^d$ with the center in zero. This ball is a $\mathbb{Z}_p$-module.

\begin{lemma}\qquad{\sl The group $O_d$ is the stabilizer of the unit ball ${\cal V}$ in the group of non degenerate linear transformations.}
\end{lemma}

\noindent{\it Proof}\qquad It is easy to see that $O_d$ is the subgroup of the stabilizer of ${\cal V}$.

Let us prove that $O_d$ coincides with the stabilizer. Assume that $g\notin O_d$ belongs to the stabilizer of ${\cal V}$. Then there exists  $x\in {\cal V}$: $gx\in {\cal V}$, $|x|_p\ne |gx|_p$. We can assume that $|x|_p< |gx|_p$ (if this will be not satisfied, i.e. we will have $|x|_p> |gx|_p$, then we can consider the element $g^{-1}$ of the stabilizer instead of $g$ and will get $|x|_p<|g^{-1}x|_p$).

Let us normalize $x$ and consider the element $y=x|x|_p$. Then $|y|_p=1$. We have by the choice of $x$ the inequality $|gy|_p>1$. Therefore $gy\notin {\cal V}$ and $g$ can not belong to the stabilizer of ${\cal V}$.

This finishes the proof of the lemma. $\square$

\begin{lemma}\label{columns}\qquad
{\sl The group $O_d$ consists of the following matrices in
$\mathbb{Q}_p^d$: a set of columns of a matrix from $O_d$ is a set of
vectors of unit norm in $\mathbb{Q}_p^d$ which generates the
$\mathbb{Z}_p$-module ${\cal V}$.

The same statement holds for the set of lines of $g\in O_d$.

%Any line of a matrix from $O_d$ has the norm one.
}
\end{lemma}

\noindent{\it Proof}\qquad The $n$-th column of a matrix of linear mapping is the image of the vector $(0,\dots,1,\dots, 0)$ with one at the $n$-th place and zeros at other places. Therefore for the matrix in $O_d$ the norm of any column is equal to one.
For a matrix $g$ the image $g{\cal V}$ is the $\mathbb{Z}_p$-module generated by columns of the matrix.

Then the first statement of the lemma follows from the previous lemma.

The module ${\cal V}'$ conjugated to ${\cal V}$ (i.e. the set of $\mathbb{Z}_p$-linear homomorphisms ${\cal V}\to\mathbb{Z}_p$) is isomorphic to ${\cal V}$. The elements of ${\cal V}'$ can be considered as lines $(k_1,\dots,k_d)$ acting on columns $(x_1,\dots,x_d)$, $k_i,x_i\in\mathbb{Z}_p$ by the $\mathbb{Z}_p$-valued scalar product:
$$
kx=\sum_{i=1}^dk_ix_i.
$$

The group $O_d$ possesses the natural right action on ${\cal V}'$ by application to vectors in ${\cal V}'$ of matrices $g'$, $g\in O_d$, where the matrix $g'$ is the transponated matrix to the matrix $g\in O_d$:
$$
(g'k)_j=\sum_{i=1}^d g_{ij}k_i.
$$

Then, since $g\in O_d$ maps the module ${\cal V}$ on itself, the same statement should be satisfied for the conjugated module ${\cal V'}$ and the transponated matrix $g'$. Therefore the lines of the matrix $g\in O_d$ (i.e. the columns of the transponated matrix) should generate the module which coincides with the unit ball with the center in zero.

This finishes the proof of the lemma.
$\square$

\bigskip

\noindent{\bf Remark}\qquad The above lemma is the $p$-adic
analogue of the proposition which states that the set of columns (lines) of an
orthogonal matrix in $\mathbb{R}^d$ is an orthonormal basis in
$\mathbb{R}^d$ and, conversely, any matrix with columns (lines) from some
orthonormal basis is orthogonal.

\begin{lemma}\label{transitivity}\qquad{\sl The group $O_d$ acts transitively on the unit sphere in $\mathbb{Q}_p^d$, i.e. for any pair of vectors
from the unit sphere there exists a transformation from $O_d$ which
maps one of the vectors to the other.}
\end{lemma}

\noindent{\it Proof}\qquad Let us choose the basis $\{e_i\}$, $i=1,\dots,d$ in ${\cal V}$ with basis vectors $e_i=(0,\dots,1,\dots,0)$ with 1 at the $i$-th place and zeros at the other places.

It is sufficient to prove the transitivity of $O_d$ for any pair $e_i$, $x$, where $e_i$ belongs to the above basis and $|x|_p=1$.

An element of the unit sphere has the form
$$
x=\left(x_1,\dots,x_d\right),\qquad x_l\in\mathbb{Z}_p,
$$
where for at least one of $x_l$ we have $|x_l|_p=1$.

Assume that $|x_1|_p=1$.

Consider the following linear mapping which maps $e_1$ to $x$:
$$
e_1\mapsto \sum_{l=1}^{d}x_le_l,\qquad e_i\mapsto e_i,\quad i\ne 1.
$$
The matrix of this mapping is obtained from the unit matrix by replacing of the first column by $x$.
Since $\{x,e_2,e_3,\dots,e_d\}$ generate the $\mathbb{Z}_p$-module ${\cal V}$, the above linear map lies in $O_d$.

This finishes the proof of the lemma.
$\square$

\bigskip

\noindent{\bf Remark}\qquad The same statement is satisfied for the right action of $O_d$ by transponated matrices.

\bigskip

Consider the group of transformations $G$ generated by matrices from
$O_d$, by arbitrary translations, and by the dilations which are homogeneous over all coordinates:
$$
x\mapsto p^{\gamma}x, \quad\gamma\in\mathbb{Z}, \quad x\in\mathbb{Q}_p^d.
$$
These transformations form three subgroups $G_1$, $G_2$, $G_3$ of $G$.

We consider the representation of the group $G$ acting in the space
$L^2(\mathbb{Q}^d_p)$ by unitary trans\-for\-ma\-tions, i.e.
matrices in $O_d$ act as
$$
f(x)\mapsto f(gx),
$$
translations act as
$$
f(x)\mapsto f(x+b),
$$
and dilations by degrees of $p$ act as
$$
f(x)\mapsto p^{-{d\gamma\over 2}} f\left(p^{\gamma}x\right).
$$

\begin{lemma}\qquad
{\sl An arbitrary element $g$ of the group $G$ can be uniquely
represented as a product of elements of the above three subgroups of $G$,
$g^{(1)}\in G_1$, $g^{(2)}\in G_2$, $g^{(3)}\in G_3$:
\begin{equation}\label{prod_of_3}
g=g^{(3)}g^{(2)}g^{(1)}.
\end{equation}
Here $g^{(1)}$ is a matrix in $O_d$, $g^{(2)}$ is a translation, and
$g^{(3)}$ is a homogeneous dilation $x\mapsto p^{\gamma}x$,
$\gamma\in\mathbb{Z}$.}
\end{lemma}

\noindent{\it Proof}\qquad This statement follows from the
observation that if we multiply $g\in G$ of the form
(\ref{prod_of_3}) from the left by the element $g'$ from any of the
above three subgroups of $G$, then the product can be put into the
form (\ref{prod_of_3}).

The uniqueness of the above representation is straightforward.
Assume we have
$$
g=g^{(3)}g^{(2)}g^{(1)}=g^{'(3)}g^{'(2)}g^{'(1)}.
$$
Then, considering the action of $g=g^{(3)}g^{(2)}g^{(1)}\in G$ in
the {\it affine} space $\mathbb{Q}^d_p$, we see that the length of
any vector in the affine space is transformed by the multiplication
by $p^{\gamma}=g^{(3)}$. Therefore $g^{(3)}$ in the above expansion
is defined uniquely by $g\in G$, i.e. $g^{(3)}=g^{'(3)}$.

Then, considering the equality $g^{(2)}g^{(1)}=g^{'(2)}g^{'(1)}$,
and taking into account that the transformation $g^{(2)}g^{(1)}$
maps $0\in \mathbb{Q}^d_p$ into  $g^{(2)}0\in \mathbb{Q}^d_p$, we
get that $g^{(2)}$ is defined uniquely by $g\in G$. This proves the
uniqueness of the representation. $\square$

\bigskip

The next theorem gives the interpretation of the $d$-dimensional basis of wavelets (\ref{basis_1}), (\ref{basis_2})  using the system
of coherent states (i.e. the orbit of the above representation) for the group $G$.

\begin{theorem}\quad{\sl The orbit of the function
$\psi^{(1)}(x)=\chi(p^{-1}x_1)\Omega(|x|_p)$, $x=(x_1,\dots,x_d)\in
\mathbb{Q}^d_p$, with respect to the defined above unitary representation of the
group $G$ is the frame in $L^2(\mathbb{Q}^d_p)$ which consists of all products of vectors of
the $p$-adic wavelet basis $\{\psi_{\gamma n J}\}$ given by (\ref{basis_1}), (\ref{basis_2}) and $p$-roots of one.}
\end{theorem}

\noindent{\it Proof}\qquad Consider the function
$\psi^{(1)}(x)=\chi(p^{-1}x_1)\Omega(|x|_p)$. This function is the
tensor product of the one dimensional wavelet $\psi(x_1)$ and the
characteristic function of the $d-1$-dimensional ball.

One can see that the orbit of the function $\psi^{(1)}$ with respect
to the subgroup $G_1=O_d$ of the $d$-dimensional
linear transformations conserving the $d$-dimensional norm is exactly the set of wavelets $\psi_J$,
defined by (\ref{basis_0}), (\ref{basis_01}). Let us consider
$$
\psi^{(1)}\left(g^{(1)}x\right)=
\chi\left(p^{-1}\sum_{n=1}^{d}g^{(1)}_{1n}x_n\right)\Omega(|x|_p).
$$
Here $g^{(1)}_{mn}$ is the matrix of the linear transformation
$g^{(1)}\in O_d$.

The set of numbers $\left(g^{(1)}_{1n}\right)$, as a vector in
$\mathbb{Q}_p^d$ lies in the unit sphere by lemma \ref{columns}.

Thus $\psi^{(1)}\left(g^{(1)}x\right)$ has the form $\psi_k(x)$ for
$k\in\mathbb{Q}_p^d$, $k=\left(g^{(1)}_{1n}\right)$, $n=1,\dots,d$.
The local constancy of the function $\psi_k(x)$ with respect to $k$
implies that $\psi^{(1)}\left(g^{(1)}x\right)=\psi_J(x)$ for
$J=\left(g^{(1)}_{1n}\,{\rm mod}\,p\right)$, $n=1,\dots,d$. The
transitivity of the action of the group $O_d$ on the unit sphere (lemma \ref{transitivity} and the remark after this lemma) implies
that the action of $g^{(1)}\in O_d$ on $\psi^{(1)}$ gives all
wavelets $\psi_J$.

An arbitrary translation can be uniquely expanded into the
composition of translation of the form (\ref{basis_2}), translation
by integers $0,1,\dots,p-1$ over any of the coordinates, and
translation with the norm less than 1. Translations of $\psi_J$ of
the form (\ref{basis_2}) give all vectors from the basis of
wavelets with $\gamma=0$, translations by the mentioned integers give
multiplications by a root of one (or identical transformation), and
translations with shorter distances give identical transformations
(due to local constancy of the functions under consideration).

Homogenous dilations by $p^{\gamma}$ give (\ref{basis_1}).

This finishes the proof of the theorem. $\square$

\bigskip

\noindent{\bf Remark}\qquad An alternative way of introducing a $d$-dimensional wavelet basis is to take the set of tensor
products of $d$ copies of the one dimensional wavelet basis. The
corresponding $d$-dimensional wavelet basis contains the vectors
$$
\otimes_{l=1}^{d}\psi_{\gamma_{l} n_{l} j_{l}},
$$
where $\psi_{\gamma_{l} n_{l} j_{l}}$ are one dimensional
wavelets (\ref{thebasis}).

This basis possesses the following interpretation as a system of
coherent states (i.e. as an orbit of a group action). The orbit of the action of the direct product of
the $d$ one dimensional affine groups applied to the tensor product
of one dimensional wavelets $\otimes_{l=1}^{d}\psi$ is the
frame consisting of the products of vectors from the above basis and
$p$-roots of one.

\bigskip

We see that the different multidimensional wavelet bases (actually
the corresponding frames which contains the products of wavelets
from the discussed bases and $p$-roots of one) can be
considered as systems of coherent states for the different groups.

This suggests the problem of investigating of the systems of
coherent states (orbits in the linear representation, cf.
\cite{Perelomov}) for the different $p$-adic groups of linear
transformations and of the construction of the corresponding bases and
frames of wavelets. This approach should be related to the
construction of matrix dilations for multidimensional wavelet bases
in the theory of real wavelets (see \cite{Daubechies}, \cite{Meyer},
\cite{Skopina} for a discussion of the matrix
dilations).

%In particular the representation theory of $p$-adic groups of
%transformations is related to the
%wavelet analysis.

\bigskip

\noindent{\bf Acknowledgments}\qquad One of the authors (S.K.) would
like to thank I.V.Volovich, V.S.Vla\-di\-mi\-rov and A.Yu.Khrennikov
for fruitful discussions and valuable comments. He gratefully
acknowledges being partially supported by the grant DFG Project 436
RUS 113/809/0-1 and DFG Project 436 RUS 113/951, by the grants of
the Russian Foundation for Basic Research  RFFI 05-01-04002-NNIO-a
and RFFI 08-01-00727-a, by the grant of the President of Russian
Federation for the support of scientific schools NSh 3224.2008.1 and
by the Program of the Department of Mathematics of the Russian
Academy of Science ''Modern problems of theoretical mathematics'',
and by the program of Ministry of Education and Science of Russia
''Development of the scientific potential of High School, years of
2009--2010'', project 3341. He is also grateful to IZKS (the
Interdisciplinary Center for Complex Systems) of the University of
Bonn for kind hospitality.

\end{document}